\documentclass[twocolumn,aps,pra]{revtex4}
\usepackage{amsmath,amssymb,mathrsfs,bm,braket,quantikz}
\usepackage[switch,columnwise]{lineno}
\usepackage{xcolor}
\usepackage{graphicx,dcolumn,times}
\usepackage{CJK}
\usepackage{bm}
\usepackage{color}
\usepackage[colorlinks,linkcolor=blue,anchorcolor=blue,citecolor=blue,filecolor=blue,menucolor=blue,runcolor=blue,urlcolor=blue,frenchlinks=blue]{hyperref}
\usepackage{lineno,hyperref}
\usepackage{changes}
\usepackage{appendix}
\usepackage{tikz,pgf}
\usepackage{tabularx}
\usepackage{multirow}

\def \H {\mathcal{H}}

\def \k {\bm{k}}

\def \H {\mathcal{H}}

\def \F {\mathcal{F}}
\def \k {\mathbf{k}}

\def \M {\mathcal{M}}

\def \R  {\mathbb{R}}

\begin{document}
     \title{Probing Tensor Singularities and Their Euler-Class Descendants via Non-Abelian Quantum Geometry Measurement}
	
	\author{Zhe Wang$^{1,5,6,}$\footnotemark[1], Yan-Qing Zhu$^{2,3}$\footnote{These authors contributed equally to this work.}$^{,}$\footnote[2]{Corresponding author. Email:zhuyq1992@gmail.com}, Xinsheng Tan$^{1,5,6,7,9,10,}$\footnotemark[1]$^,$\footnote[4]{Corresponding author. Email:tanxs@nju.edu.cn}, Giandomenico Palumbo$^{4,}$\footnote[3]{Corresponding author. Email:giandomenico.palumbo@gmail.com}, Lichang Ji$^{1,5,6}$, Wei Xin$^{1,5,6}$, Shi-Liang Zhu$^{2,3,8,9,}$\footnote[5]{Corresponding author. Email:slzhu@scnu.edu.cn}, Yang Yu$^{1,5,6,7,9,10,}$\footnote[6]{Corresponding author. Email:yuyang@nju.edu.cn}}  

	\affiliation{$^1$National Laboratory of Solid State Microstructures, School of Physics, Nanjing University, Nanjing 210093, China}	
	\affiliation{$^2$Key Laboratory of Atomic and Subatomic Structure and Quantum Control (Ministry of Education), Guangdong Basic Research Center of Excellence for Structure and Fundamental Interactions of Matter, School of Physics, South China Normal University, Guangzhou 510006, China}
	\affiliation{$^3$Guangdong Provincial Key Laboratory of Quantum Engineering and Quantum Materials,
		Guangdong-Hong Kong Joint Laboratory of Quantum Matter,
		Frontier Research Institute for Physics, South China Normal University, Guangzhou 510006, China}
	\affiliation{$^4$CFisUC, Department of Physics, University of Coimbra, Rua Larga, 3004-516 Coimbra, Portugal}
	\affiliation{$^5$Shishan Laboratory, Suzhou Campus of Nanjing University, Suzhou 215000, China}
	\affiliation{$^6$Jiangsu Key Laboratory of Quantum Information Science and Technology, Nanjing University, Suzhou 215163, China}
	\affiliation{$^7$Synergetic Innovation Center of Quantum Information and Quantum Physics, University of Science and Technology of China, Hefei, Anhui 230026, China}
	\affiliation{$^8$Quantum Science Center of Guangdong-Hong Kong-Macao Greater Bay Area, 3 Binlang Road, Shenzhen, China}
	\affiliation{$^9$Hefei National Laboratory, Hefei 230088, China}
	\affiliation{$^{10}$Jiangsu Physical Science Research Center, China}


	\begin{abstract}
        We report  the theoretical prediction and experimental observation of a new class of four-dimensional (4D) tensor singularities and their three-dimensional (3D) Euler-class descendants, protected by chiral and spacetime inversion symmetries on a superconducting circuit platform. The 4D point-like singularity/monopole, characterized by the Dixmier-Douady class of a real bundle gerbe associated with tensor gauge fields, is observed to evolve into a nodal ring carrying an additional first Euler class charge under symmetry-preserving perturbations. Dimensional reduction reveals 3D Euler and Euler curvature dipoles, exhibiting nontrivial Euler topology and a topological sum rule that ensures zero-energy flat bands inherit nontrivial topology even without interactions. Crucially, these high-dimensional degenerate systems are mapped and reconstructed using a hybrid analog-digital protocol designed for non-Abelian quantum geometry measurement within a superconducting qubit array. Our work not only expands the family of topological monopoles but also establishes a robust experimental framework for exploring high-order gauge theory and real-bundle topology across diverse quantum platforms.
	\end{abstract}
		\maketitle
	
\noindent\emph{\color{blue}Introduction.}---
Magnetic monopoles, from Dirac’s 1931 $U(1)$ proposal \cite{Dirac1931} to Yang’s non-Abelian generalizations \cite{CNYang1978,Hooft1974,Polyakov1974}, remain central to restoring fundamental symmetries and explaining charge quantization. Beyond the Standard Model, M-theory predicts extended topological structures, such as five-branes and self-dual strings, governed by Kalb-Ramond tensor gauge fields and bundle gerbes \cite{Howe1998,Saemann2011}. While these higher-dimensional structures are inaccessible in 4D spacetime, artificial quantum systems \cite{DWZhang2018,Cooper2019,Ozawa2019,HXue2022,Sahin2025} provide a versatile platform to explore their physics. Guided by this potential, research has branched higher-spin \cite{Bradlyn2016,YQZhu2017} and higher-dimensional systems \cite{SCZhang2001,Price2015,Zilberberg2018,YQZhu2022,Bouhiron2024,Bouhon2024}, enabling the realization of Dirac \cite{Tarruell2012}, Weyl \cite{SYXu2015,ZYWang2021,Tan2019}, Maxwell\cite{XSTan2018}, and Yang monopoles \cite{SSugawa2018}, alongside Abelian 4D tensor monopoles (TMs) \cite{Palumbo2018,YQZhu2020,XSTan2021,MChen2022,YZhang2024,QMo2025}.These TMs, classified by the Dixmier-Douady (DD) invariant \cite{Palumbo2019}, were detected through Abelian quantum metric measurements \cite{MYu2019,XSTan2019b}. However, it remains unexplored whether non-Abelian band degeneracies can host SO(2) tensor singularities and how such higher-dimensional objects dictate unconventional topology--like Euler-class characteristics \cite{Ahn2019,Bouhon2020}--in lower-dimensional subsystems.

In this Letter, we address these challenges by combining theoretical prediction with experimental implementation in a superconducting qubit array. We report a new class of SO(2) tensor singularities and their 3D topological descendants, both protected by chiral and spacetime-inversion symmetries. These 4D point-like TMs are characterized by topological invariants linked to the real DD class of a real bundle gerbe \cite{Hekmati2019,Jankowski2025}. We demonstrate that symmetry-preserving perturbations inflate these singularities into nodal-ring structures carrying an additional topological invariant associated with the first Euler class. Furthermore, dimensional reduction reveals novel 3D counterparts--Euler dipoles and Euler curvature dipoles--exhibiting nontrivial Euler topology \cite{Ahn2019,Bouhon2020,Jankowski2024}. Crucially, we find that the first Euler numbers satisfy a topological sum rule and that perfectly zero-energy flat bands inherit nontrivial topology even without interactions \cite{Peri2021,Kwon2024,Wahl2025}. By engineering a parametrized manifold in superconducting circuits, we measure the non-Abelian quantum geometry via adiabatic evolution to characterize these 4D and 3D singularities. Our results expand the landscape of topological phenomena and establish a scalable pathway for exploring fundamental monopoles and higher-order gauge theories in quantum simulators.

\begin{figure}[http]
		\centering
		\includegraphics[width=8.7cm]{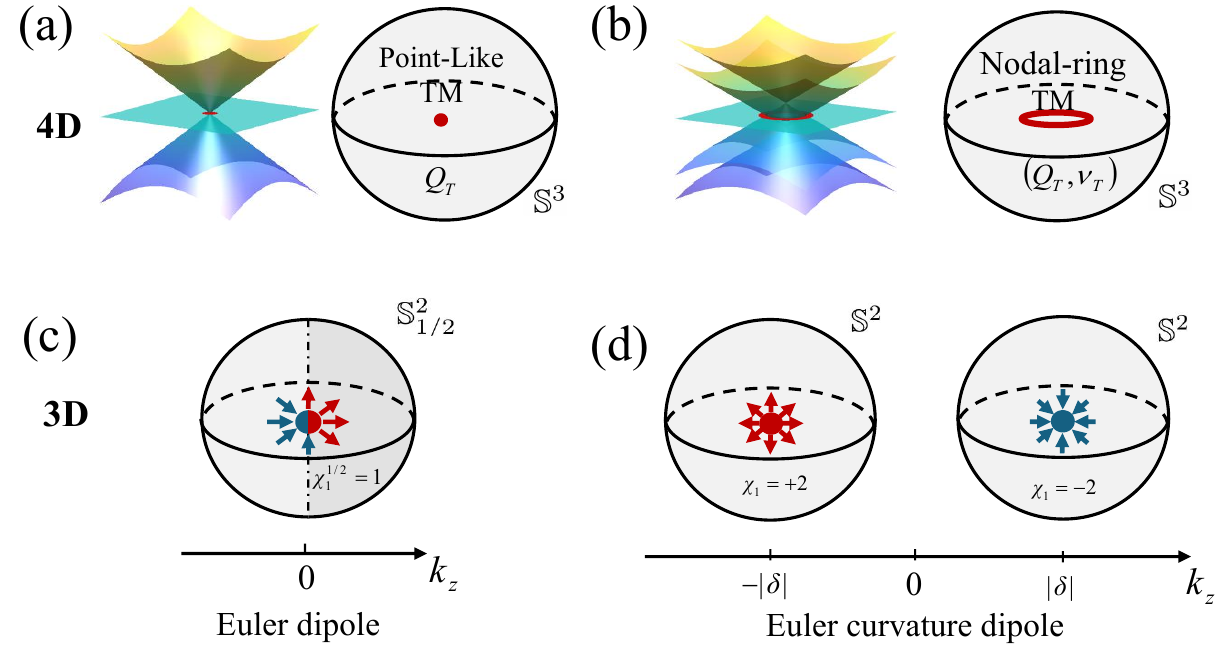}
		\caption{Schematic of 4D SO(2) TMs and their 3D descandants. (a) Point-like TM  at $\delta=0$ with spectrum $(E,\mathbf{k}_{\|})$ for $\mathbf{k_{\perp}}=0$,  characterized by charge $Q_T=1$. 
		    (b) Nodal-ring inflated monopole for $\mathbf{k_{\perp}}=0$, $\delta\neq 0$, where the band-crossing points  form a nodal ring in $\mathbf{k}_{\|}$ plane, carrying dual topological charges $(Q_T,\nu_T)=(1,2)$. 
			   (c) Euler dipole located for $\delta=0$, where two oppositely charged Euler monopoles coincide in the origin. (d) Euler curvature dipole for 
			$\delta\neq 0$, with the monopoles displaced along $k_z$ axis.}\label{fig1}
	\end{figure}

	\noindent 
	\emph{\color{blue}Tensor singularities and their 3D descendants.}---
	We begin with a minimal model in 4D space, $\mathbf{k} = (k_x, k_y, k_z, k_w)$, described by a six-band Hamiltonian,
	\label{H}
	\begin{equation}\label{Model}	\H_1(\mathbf{k})=k_x\lambda_2\otimes\sigma_2+k_y\lambda_1\otimes\sigma_0+k_z\lambda_4\otimes\sigma_3+k_w\lambda_4\otimes\sigma_1,
	\end{equation}
	where $\sigma_i$ are the Pauli matrices ($\sigma_0$ is the $2\times2$ identity) and $\lambda_i$ denote selected $3\times3$ Gell-Mann matrices \cite{Gellman}.
	This Hamiltonian is real and respects both spacetime-inversion ($PT$) and chiral ($S$) symmetries:
	\begin{equation}
		[PT, \mathcal{H}_1(\mathbf{k})] = 0, \quad \{S, \mathcal{H}_1(\mathbf{k})\} = 0,
	\end{equation}
	where $PT = \mathcal{K}$ satisfying $(PT)^2 = +1$, with $\mathcal{K}$ denoting complex conjugation. The chiral operator $S = \text{diag}(-1, 1, 1) \otimes \sigma_0$ satisfies $S^2 = +1$. Here, $P$ and $T$ denote inversion and time-reversal symmetries, respectively. 
	As illustrated in Fig.~\ref{fig1}(a), the energy spectrum $E_n(\mathbf{k}) = n|\mathbf{k}|$ ($n=0, \pm1$) features twofold degenerate bands, forming a sixfold band crossing at the origin. This singularity realizes a new class of TM characterized by an SO(2) gauge structure. Specifically, the associated 3-form non-Abelian tensor Berry curvature $H_{\mu\nu\lambda}$ \cite{Palumbo2021} for each degenerate eigenspace is a $2\times2$ antisymmetric matrix in the real gauge \cite{NATB}. Thus,  the conventional DD invariant \cite{Palumbo2018,YQZhu2020}, for SO(2) TMs vanishes here, $\mathcal{DD} = \frac{1}{2\pi^2}\int \text{tr}\,H_{\mu\nu\lambda} = 0$.
	
	\bigskip

Instead, we define the topological charge of this monopole as
\begin{equation}
	Q_T = \frac{1}{2\pi^2} \int_{\mathbb{S}^3} d\mathbf{S}\cdot \mathbf{H},
\end{equation}
where $d\mathbf{S}$ is the oriented area element on the enclosing 3-sphere $\mathbb{S}^3$,  $\mathbf{H}=(H^x,H^y,H^z,H^w)$ and $H^{\rho}=\frac{1}{2\times 3!}\epsilon^{\rho\mu\nu\lambda}\text{tr}(\Phi H_{\mu\nu\lambda})$ with $\epsilon^{\rho\mu\nu\lambda}$ is the Levi-Civita symbol. 
Here $\Phi$ denotes the momentum-space Higgs field, constrcuted as the sewing matrix of mirror symmetry  along the extended dimension (acting on the Hamlitonians $[\M,\H_1]=0$) in twofold degenerate eigenspace,  which explictly takes the form $\Phi = \sigma_2$ \cite{SM}. Note that the topological classification of this SO(2) TM is connected to the real Dixmier-Douady class \cite{Hekmati2019,Jankowski2025}, and the system falls into the $\mathbb{Z}$-class~\cite{JXDai2024}.
Direct calculation shows that only the dispersive bands carry nontrivial topology ($Q_T = 1$ for $n = \pm 1$) with  $\mathbf{H}=\frac{\mathbf{k}}{k^4}$ serving as the source of the 4D point-like monopole while the middle flat bands are topologically trivial ($Q_T = 0$) with $\mathbf{H}=0$.

Introducing a symmetry-preserving perturbation $\Delta=\delta \lambda_4 \otimes \sigma_0$ splits each degenerate band into $E_n^{\pm} = n \sqrt{ k_{\perp}^2 + ( k_{\|} \pm \delta )^2 }$ ($n=0, \pm 1$), transforming the point-like TM into a nodal ring of radius $|\delta|$ in the $\mathbf{k}_{\|}$ plane [Fig. \ref{fig1}(b)]. Here $k_{\|}=\sqrt{k_z^2+k_w^2}$ and $k_{\perp}=\sqrt{k_x^2+k_y^2}$.
This \emph{inflated monopole} now carries two distinct topological charges: the conserved tensor monopole charge $Q_T$, and a second invariant $\nu_T=\chi_1^{\text o}-\chi_1^{\text i}$, with the first Euler number $\chi_1$ of the gapped 2D subsystems $\H_{1}^{\mathbf{k}_{\|}}(\mathbf{k}_{\perp})$, computed as, 
\begin{equation}
	\begin{split}
		\chi_1=\frac{1}{2\pi}\int_{\R^2} d\mathbf{S}\cdot\mathbf{F}_\text{E},\\
	\end{split}\label{Euler}
\end{equation}
where $d\mathbf{S}$ is the oriented area element on the entire 2D flat space $\R^2$ spanned by $\mathbf{k}_{\perp}=(k_x,k_y)$,  $\mathbf{F}_\text{E}=(F^x_\text{E},F^y_\text{E},F^z_\text{E})$ and the component of Euler curvature vector $F^{\rho}_\text{E}=\frac{1}{4 i}\epsilon^{\rho\mu\nu}\text{tr}(\Phi\F_{\mu\nu})$ with $\epsilon^{\rho\mu\nu}$ is the Levi-Civita symbol. Explictly, $\chi_1=\frac{1}{2\pi}\int_{\mathbf{k}_{\perp}}dk_xdk_y \F_{xy}^{12}$, where $\mathcal{F}_{xy}^{12}$ is the off-diagonal element of the real antisymmetric $2\times 2$ matrix $\mathcal{F}_{xy}$. 
Here $\mathbf{k}_{\|} = (k_z, k_w)$ is held fixed, with the radial distance $k_{\|}$ in the $k_z$-$k_w$ plane. The quantities $\chi_1^{\text{i}}$ and $\chi_1^{\text{o}}$ denote the first Euler numbers evaluated for $\mathcal{H}_1^{\mathbf{k}_{\|}}(\mathbf{k}_{\perp})$  inside the nodal ring ($k_{\|} < |\delta|$) and  outside the nodal ring ($k_{\|} > |\delta|$), respectively.
Direct calculations reveal that the topological invariants $\chi_1^n$, defined for each twofold subspace, satisfy the sum rule $\sum_n \chi_1^n = 0$ with $\chi_1^n = \frac{1}{2}\kappa_n$. Specifically, $(\kappa_{\pm1}, \kappa_0)$ shifts from $(1, -2)$ for $k_{\|} < |\delta|$ to $(-1, 2)$ for $k_{\|} > |\delta|$ indicating the radial phase transition \cite{SM}. Crucially, we define a second topological invariant $\nu_T = \chi_1^{\text{o}} - \chi_1^{\text{i}} = 2$ solely from the $2\mathbb{Z}$ Euler topology \cite{JXDai2024} of the exactly zero-energy flat bands ($n=0$). This jump in the flat-band Euler number characterizes the nodal-ring tensor monopole and identifies it as a source of 3D monopoles, as discussed below.

By setting $k_w = 0$ for simplicity, $\mathcal{H}_1$ is reduced to a 3D system where the mirror symmetry $\mathcal{M}_z$ is restored \cite{Mz}. At $\delta=0$, the band crossing at $k_z=0$ realizes an Euler dipole--an SO(2) generalization of Berry dipoles \cite{Graf2023,ZYZhuang2024} where two oppositely charged Euler monopoles coincide. Its curvature vector mimics a magnetic dipole: $\mathbf{F}_\text{E}=\kappa_n(\mathbf{d}_0\cdot\mathbf{k})\mathbf{k}/k^4$, with $\mathbf{d}_0 = (0, 0, 1)$ and $\kappa_{\pm1}=-1, \kappa_0=2$. While the total Euler number over $\mathbb{S}^2$ vanishes, the dipole nature is captured by the half-sphere $\mathbb{S}^{1/2}$ integration $\chi_1^{k_z>0} = -\chi_1^{k_z<0} = \tfrac{1}{2}\kappa_n$. We use the flat-band value $\chi_1^{1/2}=1$ to characterize this configuration [Fig. \ref{fig1}(c)]. Tuning $\delta \neq 0$ separates these coincident charges into a pair of Euler monopoles at $\mathbf{k}_{\pm} = (0,0,\pm |\delta|)$ [Fig. \ref{fig1}(d)]. Each monopole carries a flat-band Euler charge $\chi_1^{\pm} = \pm 2$, obtained by integrating over a closed sphere enclosing $\mathbf{k}_{\pm}$.
	
\bigskip

\noindent\emph{\color{blue}Hybrid simulating protocol via adiabatic procedure.}---
To realize these theoretical predictions, we propose an unified protocol for measuring quantum geometry via adiabatic evolution, including both the real and imaginary components of the Abelian and non-Abelian forms. Previous schemes faced substantial implementation challenges, including extremely weak signals and a lack of generality. Our hybrid analog–digital approach enables direct access to this quantity, which proceeds in three steps: 1. Construction of the adiabatic evolution path with Hamiltonian $H(t)$.
2. Enhancement of the signal-to-noise ratio using the so-called mTQDA (modified transitionless quantum driving) technique\cite{YZhang2024}, while the Hamiltonian is modified as $H'(t)=H(t)+\lambda H_{cd}$, here $H_{cd}$ and $\lambda$ are the counterdiabatic Hamiltonian and its factor \cite{supp}.
3. Decomposition of the full adiabatic operation into a sequence of circuit gates.
The entire adiabatic evolution operator $U$ is defined as
\begin{equation}
	U = \mathcal{T} \exp\left[-i\int { R} H'(t) R^{-1}\,\mathrm{d}t \right],
	\label{U}
\end{equation}
where $R$ rotates the initial state onto a single-qubit excited state and $\mathcal{T}$ denotes time-ordering \cite{SM}. 



This hybrid approach enhances the measurable response by two to three orders of magnitude and enables systematic parameter calibration, which overcomes the intrinsic limitations of conventional protocols, and it completes the measurement process in a controlled and scalable manner \cite{SM}.
We verified this experimentally in a superconducting circuit consisting of six transmon qubits arranged in a $2\times3$ lattice. Nearest-neighbour interactions are mediated by tunable couplers with parametric modulation. The six-dimensional Hilbert space is defined within the single-excitation manifold spanned by the first excited states of the qubits \cite{SM}.  


\begin{figure*}[t]
	\includegraphics[width=1.7\columnwidth]{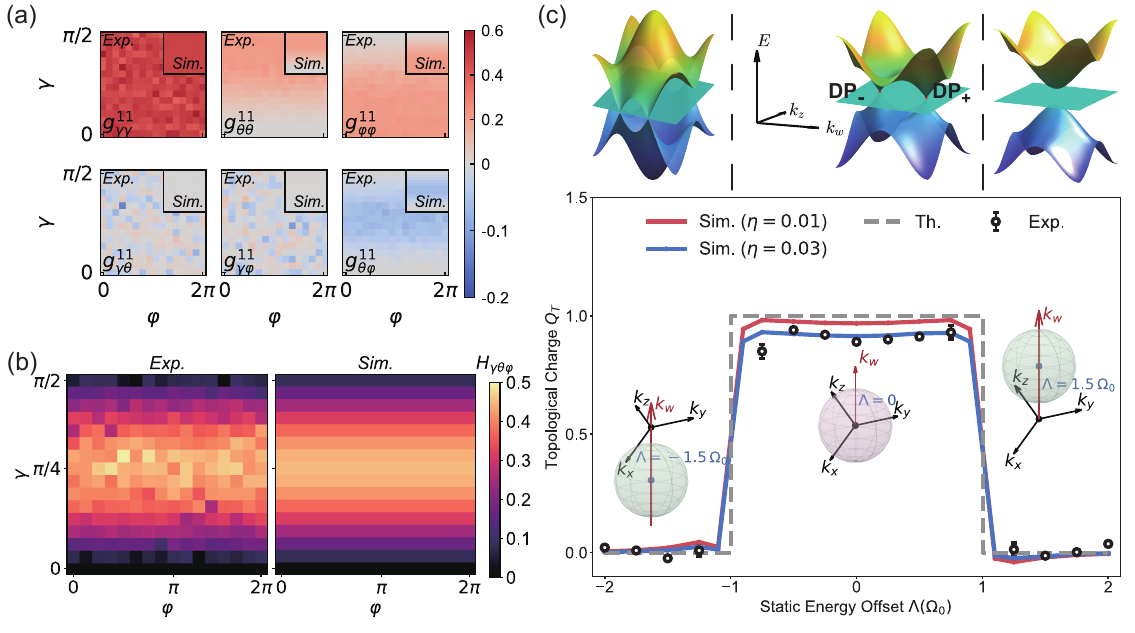}%
	\caption{
		\textbf{Topological characterization of the SO(2) TM.}
\textbf{(a)} Measured non-Abelian quantum metric components vs numerical simulations.
\textbf{(b)} Reconstructed tensor Berry curvature $H_{\gamma\theta\phi}$; two independent directions are shown due to symmetry.
\textbf{(c)} Topological invariant $Q_T$ for closed hypersurfaces surrounding (center) or avoiding (left/right) the monopole. Experimental data (black circles) are compared with numerical simulations at excitation densities $n_{\mathrm{ex}}=0.01$ (red) and $0.03$ (blue), representing different “amplifier gains”. Insets: integration surfaces. Top: Tight-binding bands in the $(k_z,k_w)$ plane at $k_x=k_y=0$ for varied bias $\Lambda$.
	}
	\label{fig:topology}
	\label{fig:metric_results}
\end{figure*}

\begin{figure*}[t]
	\includegraphics[width=1.7\columnwidth]{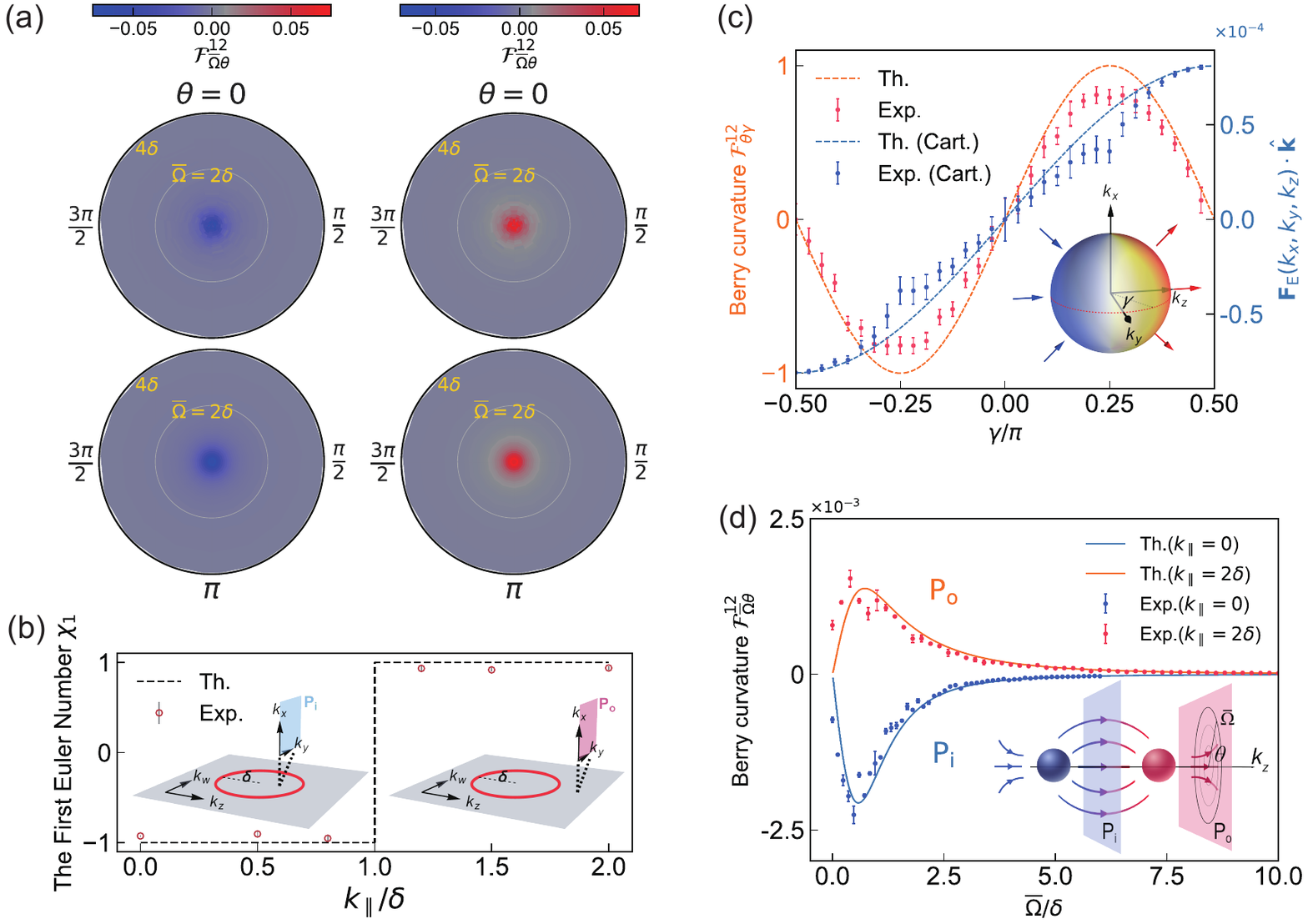}%
	\caption{
		\textbf{(a)} Nodal ring topology. Top: Measured curvature distributions on momentum slices with 
		$k_\parallel = 0.8|\delta|$ (inside the ring) and $1.2|\delta|$ (outside), respectively.
		Bottom: Numerical simulations. 
		\textbf{(b)} Extracted Euler number $\chi_1$ versus $k_\parallel$. 
		The experimental results(red dots) and theoretical prediction(black dashed line) are matched well. 
		The jump from $\chi_1=+1$ to $-1$ across the nodal ring confirms the nodal-ring invariant $\nu_T=2$. Insets: schematic illustration of the spatial location of the integration plane.
		\textbf{(c)} Measured $\mathcal{F}_{\theta\gamma}^{12}$ in 3D Euler Dipole, theoretical value(bright orange dashed line) and experimental data(red circles with error bars) show well agreement. The Berry curvature is also shown in Cartesian coordinates for comparison with the theoretical analysis. Inset: Schematic of the chosen meridian in the 3D parameter space.
		\textbf{(d)} The Berry curvature $\mathcal{F}_{\bar{\Omega}\theta}^{12}$ in the $k_\perp$ plane at $k_\parallel = 0$ (blue) and $k_\parallel = 2\delta$ (red), showing theory (curves) and experimental data (circles with error bars). Insets: Schematic of the integration $k_x$-$k_y$ plane within the 3D parameter space.
	}
	\label{fig:euler_ring}
\end{figure*}

\noindent\emph{\color{blue}4D Point-like TM and the non-Abelian quantum metric.---}
In the experiment, $\H_1(\k)$ presented in Eq.\eqref{Model} is reparameterized using 4D hyperspherical coordinates $(\Omega,\,\gamma,\,\theta,\,\phi)$ as:
	$k_w = -\Omega \sin\gamma \sin\phi + \Lambda, k_x = -\Omega \cos\gamma \sin\theta,
	k_y= \Omega \cos\gamma \cos\theta, k_z = \Omega \sin\gamma \cos\phi$,
where $\Lambda$ represents a static energy offset, corresponding to the topological phase transition.  We characterize the tensor topology by measuring the non-Abelian quantum metric, exploiting its intrinsic connection to the tensor Berry curvature\cite{NATM}. By performing parametric ramps along individual and pairwise directions $\{\gamma, \theta, \phi\}$, we extract all components of the geometric tensor $g_{\mu\nu}^{nn}$ within the dispersive two-fold degenerate subspace from the resulting excitation populations.

Fig.~\ref{fig:metric_results}(a) presents the experimentally reconstructed components of the non-Abelian quantum metric on representative manifolds in parameter space, which agree well with numerical simulations shown in the insets. Due to the intrinsic symmetries of the system, only six independent components are experimentally accessible, and the data shown correspond to the degenerate subspace component $g^{11}_{\mu\nu}$. 
From the measured $g_{\mu\nu}$ we extract the tenser Berry curvature, which is independent of $\theta$ due to rotational symmetry.   
Fig.~\ref{fig:metric_results}(b) compares the experimentally reconstructed $H_{\gamma\theta\phi}$ with numerical simulations.

The topological charge $Q_T$ can be now extracted by a hypersurface $\mathbb{S}^3$ enclosing the 4D origin\cite{SM},
\begin{equation}
	Q_{T}=\frac{1}{\pi}\int_0^{\pi/2} \int_0^{2\pi}d \gamma d\phi \, \sqrt{2\,\det G_\Omega}
\end{equation}
where $G_{\Omega}=\mathrm{tr}\,g_{\mu\nu}$ ($\mu,\nu=\gamma,\theta,\phi$). When no bias is applied ($\Lambda=0$), the integration hypersurfaces fully enclose the monopole, yielding a nearly quantized topological charge of $Q_T=0.88(7)$, in good agreement with the theoretical value of $1$. As $\Lambda$ increases, the hypersurfaces shift such that they eventually cease to enclose the monopole. In this regime, the extracted charge $Q_T$ switches to zero, reflecting the phenomenon of topological phase transition, as shown in Fig.~\ref{fig:topology}(c). We summarizes the experimental extracted $Q_T$ (black circles) as a function of the surface displacement $\Lambda$, alongside numerical simulations corresponding to excitation densities $n_{\mathrm{ex}}=0.01$ (red) and $0.03$ (blue), which correspond to the different "amplifier gains". We optimize the trade-off between signal-to-noise ratio and experimental error to achieve optimal performance. The gray dashed line denotes the theoretical values, demonstrating a clear topological transition.

 A complementary viewpoint is offered by the tight-binding model, in which the linear modulation of $\k$ in Hamiltonian $\H_1(\k)$ is replaced by
$
\{k_x\rightarrow\sin k_x,k_y\rightarrow \sin k_y, k_z\rightarrow\sin k_z, k_w\rightarrow \Lambda+3-\cos k_x-\cos k_y-\cos k_z-\cos k_w\,\}.
$
For $|\Lambda|<1$, the spectrum hosts two inversion-related degeneracy points forming a monopole–antimonopole pair with opposite topological charges, reflecting the system’s nontrivial topology. 
As $|\Lambda|$ approaches 1, the nodal pair annihilates, yielding a gapped spectrum for $|\Lambda|>1$ at
$k_z=0$.  
The creation and annihilation of this nodal pair provide a direct spectroscopic signature of the topological phase transition inferred from the measured $Q_T$.



\noindent\emph{\color{blue}Topology of the inflated TM and nodal-ring invariant.---}
Under a symmetry-preserving perturbation $\Delta$, the point-like tensor singularity inflates into a nodal ring in the $(k_z,k_w)$ plane. The topological properties of the nodal-ring TM are further characterized by an additional topological invariant related to the first Euler number in Eq. (\ref{Euler}), which is extracted from the off-diagonal component of the SO(2) Berry curvature $\mathcal{F}_{\nu\mu}^{12}$.  




To extract the first Euler number, we measure the Berry curvature of the $(k_x,k_y)$ subspace at varying radial positions $k_\parallel=\sqrt{k_z^2+k_w^2}$. 
Under the adopted parametrization, the two-dimensional degenerate manifold is described in generalized polar coordinates $(\bar{\Omega},\theta)$, where $\bar{\Omega}=|\Omega\cos\theta|$ denotes the effective radial parameter. 
Fig.~\ref{fig:euler_ring}(a) shows the reconstructed Berry curvature $\mathcal{F}^{12}_{\bar{\Omega} \theta}$ on representative $k_\perp$ slices.
A sharp sign reversal occurs across the nodal ring: the curvature inverts from inside ($k_\parallel < |\delta|$) to outside ($k_\parallel > |\delta|$). The observed circular color patterns confirm that the experimental data align well with the simulations.

From the curvature data, the first Euler number $\chi_1(k_\parallel)$ is obtained for each $k_\parallel$, quantifying the topology of the $\mathrm{SO}(2)$ bundle associated with the degenerate subspace,
\begin{equation}
	\chi_1(k_\parallel)
	=\int_0^{\bar{\Omega}_{max}}\int_0^{2\pi}
	\mathrm{d}\bar{\Omega}\mathrm{d}\theta\, \mathcal{F}_{\bar{\Omega}\theta}^{12}.
\end{equation}
The six experimental points shown in Fig.~\ref{fig:euler_ring}(b) correspond to 
$k_\parallel = (0, 0.5, 0.8, 1.2, 1.5, 2)\,\delta$, with the corresponding maximum radial parameter 
$\bar{\Omega}_{\mathrm{max}} = (6, 6, 4, 4, 6, 10)\,\delta$.
Data points inside the nodal ring yield $\chi_1 = -0.92(0)$, while those outside give $\chi_1 = 0.92(3)$, in quantitative agreement with the theoretical values of $\pm1$. 
This sign reversal is consistent with the theoretically predicted topological invariant. 
The difference between the two regimes defines an emergent nodal-ring invariant,
	$\nu_T=\chi_1^{\mathrm{o}}-\chi_1^{\mathrm{i}}=1.84(3)$,
in close agreement with the theoretical value of $2$. 
This quantized response reflects the underlying topology of the SO(2) TM encoded in the nodal-ring structure.

\noindent\emph{\color{blue}3D Euler dipoles and Euler curvature dipoles.---}
Our hybrid analog-digital approach exhibits strong universality, enabling the investigation of topological properties in systems with arbitrary dimensions. Experimentally, we reduce the dimension by setting 
$k_w=0$ , constructing the Hamiltonian of 3D Euler dipoles. The measurement schemes are identical to those employed for the 4D parameter space.

Here we consider two scenarios. For a point-like Euler dipole, we choose a spherical surface in the parameter space that encloses the dipole and measure the Berry curvature distributed on this surface, as shown in Fig.~\ref{fig:euler_ring}(c). Due to the axial symmetry of the system, we present the measured Berry curvature as a function of $\gamma$. It can be seen that the values on the two hemispheres of the parameter sphere are equal in magnitude but opposite in sign. Consequently, calculating the Euler characteristic over the entire surface yields a result of -0.036, in close agreement with the theoretical value of $0$.
When perturbations are introduced, the original point-like dipole splits into a pair of positive and negative charges, named Euler curvature dipole. In this case, we place two planes labeled as $P_i$ and $P_o$ in the parameter space and measure the corresponding non-Abelian Berry curvature respectively, as shown in Fig.~\ref{fig:euler_ring}(d). Our data accurately capture the spatial distribution of the effective field strength. By integrating the curvature over $P_i$ and $P_o$, we obtain approximately half of the topological charge associated with the two monopoles, yielding values of -0.92(6) and 0.93(8), respectively.

\noindent\emph{\color{blue}Discussion and outlook.}---
Our work establishes a unified framework for SO(2) tensor monopoles and their Euler-class descendants, revealing that 4D topological structures can be systematically understood via their lower-dimensional projections. The experimental measurement of the non-Abelian quantum metric and Berry curvature in a superconducting simulator overcomes the challenge of probing topological charges in degenerate manifolds, setting a new standard for resolving non-Abelian quantum geometry. These findings provide a new paradigm for bridging condensed matter and high-energy physics. The confirmed stability of these structures invites future exploration into interaction effects, non-Hermitian extensions \cite{ZGong2018,Kawabata2019,YQZhu2021,He2023}, and higher-order topological phases \cite{Benalcazar2017a,Schindler2018b,ZWang2019,YQZhu2025}. Most profoundly, the capability to simulate 4D tensor gauge fields opens a crucial pathway for quantum simulators to address fundamental questions regarding compactified dimensions and the topological nature of the early Universe.

\noindent\emph{\color{blue}Acknowledgments.---}
This work is supported by the National Key R\&D Program of China (Grants No. 2022YFA1405304, No. 2024YFA1409300), the National Natural Science Foundation of China (Grants No. 12504588, No.  U21A20436 and No. 12074179), and Quantum Science and Technology-National Science and Technology Major Project (Grant No. 2021ZD0301702), the NSF of Jiangsu Province (Grants No. BE2021015-1, No. BK20232002, and BK20233001), the NSF of Shandong Province (Grant No. ZR2023LZH002), and Open Fund of Key Laboratory of Atomic and Subatomic
Structure and Quantum Control (Ministry of Education). This work was financed through national funds by FCT - Fundação para a Ciência e Tecnologia, I.P. in the framework of the project UID/04564/2025, with DOI identifier 10.54499/UID/04564/2025.

The data generated and analyzed during this study are available from the corresponding author upon reasonable request.


\bibliographystyle{apsrev4-1}
\bibliography{myref,ref2}



\widetext
\clearpage

\end{document}